\documentclass[vecphys]{svmult}
\usepackage{amsmath}
\usepackage{amssymb}

\usepackage{makeidx}         
\usepackage{graphicx}        
\usepackage{multicol}        
\usepackage{cite}            
\usepackage[hypertex]{hyperref}                             
\usepackage{srctex}
\usepackage[bottom]{footmisc}

\DeclareMathOperator{\Tr}{Tr} \DeclareMathOperator{\Imag}{Im}
\DeclareMathOperator{\Real}{Re}

\renewcommand{\Re}{\mathop\mathrm{Re}\nolimits}
\renewcommand{\Im}{\mathop\mathrm{Im}\nolimits}

\makeindex             

\begin{document}

\title{Synchronized Andreev Transmission in Chains of SNS Junctions}
\author{N.\,M.\,Chtchelkatchev$^{1,2,3}$ \and T.\,I.\,Baturina$^{3,4}$ \and A. Glatz$^{3}$ \and V.\,M.\,Vinokur$^{3}$ }
\institute{Institute for High Pressure Physics, Russian Academy of Sciences,
Troitsk 142190, Moscow region, Russia 
\and L.D. Landau Institute for Theoretical Physics, Russian Academy of Sciences, Akademika Semenova av. 1-A,  Chernogolovka 142432, Moscow Region, Russia
\and Materials Science Division, Argonne National Laboratory, Argonne, Illinois 60439, USA
%
\texttt{vinokour@anl.gov}
\and Institute of Semiconductor Physics, 13 Lavrentjev Avenue, Novosibirsk 630090, Russia}


\maketitle

\begin{abstract}
We construct a nonequilibrium theory for the charge transfer through a diffusive
array of alternating normal (N)
and superconducting (S) islands comprising an SNSNS junction, with the size
of the central S-island being smaller than  the energy relaxation length.
We demonstrate that in the nonequilibrium regime the central island
acts as Andreev retransmitter with the Andreev conversions at both NS interfaces
of the central island correlated via over-the-gap transmission and Andreev reflection.
This results in a synchronized transmission at certain resonant voltages which can be
experimentally observed as a sequence of spikes in the differential conductivity.
\end{abstract}

\noindent

\section{Introduction}\label{Sec:Intro}
\begin{figure}[tp]
\begin{center}
\includegraphics[width=\textwidth]{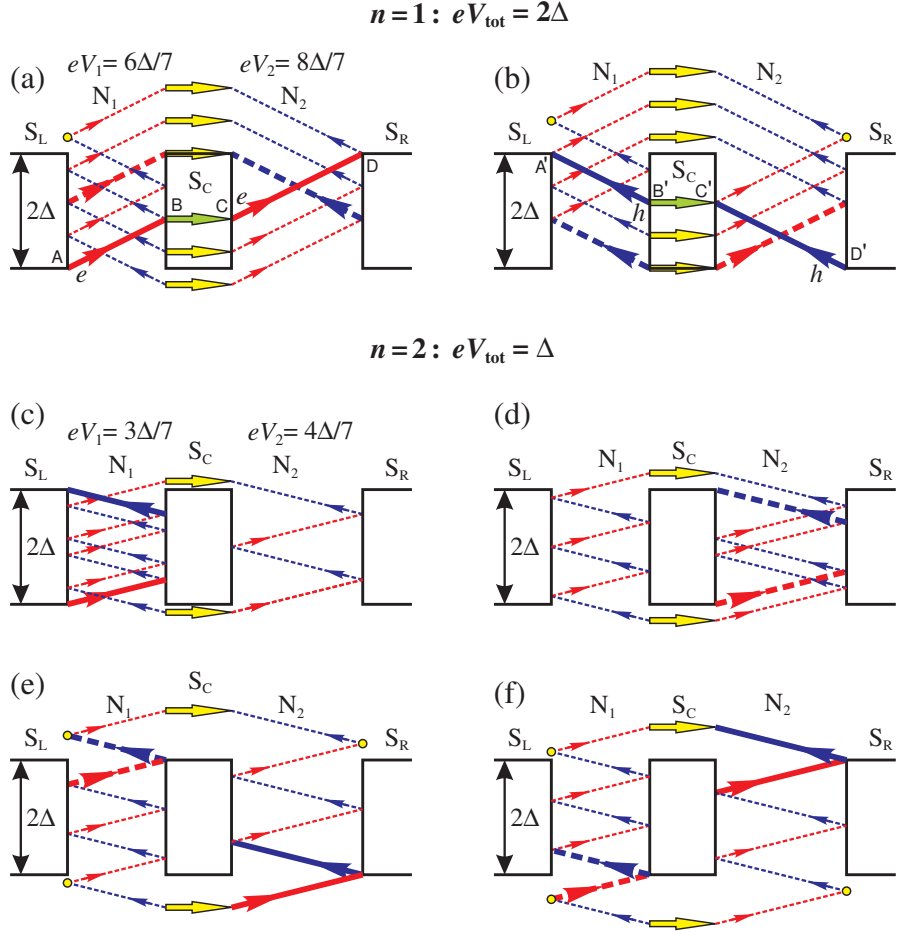}
\caption{Diagrams of the SAT processes for the first, $n=1$, (a)-(b), and second, $n=2$
(c)-(f), subharmonics of the resonant singularities in $dI/dV$ described by
Eq.\,\eqref{eq:V_position} for the SNSNS junction with the normal resistances ratio $R_1/R_2=3/4$
(depicted through 3/4 ratio of the respective lengths of the normal regions).
The thick solid lines represent the quasiparticle paths starting and/or ending at
the points of singularity in the density of states at energies
$\varepsilon=\pm\Delta$ at the electrodes S$_{\scriptscriptstyle{\mathrm L}}$
and S$_{\scriptscriptstyle{\mathrm R}}$.
The dashed solid lines show paths starting and/or ending at the
edges of the gap of the central island S$_{\scriptscriptstyle{\mathrm C}}$.
The circle denotes the over-the-gap Andreev reflections at the electrodes.
The paths ABCD [panel (a)]  and D$^\prime$C$^\prime$B$^\prime$A$^\prime$ [panel (b)]
correspond to the electron- and hole trajectories, respectively.
Synchronization of the energies of the incident and emitted quasiparticles
at points B and C (B$^\prime$ and C$^\prime$) is shown by arrows.
SAT is realized by trajectories passing through the singular points $\varepsilon=\pm\Delta$
of the central island S$_{\scriptscriptstyle{\mathrm C}}$
and including over-the-gap transmissions and Andreev reflections.
Trajectories synchronizing other transmissions across S$_{\scriptscriptstyle{\mathrm C}}$
and those of higher orders are not shown.
Note, that voltage drops $eV_1=6\Delta/7$ and $eV_2=8\Delta/7$ [panels (a)-(b)]
($eV_1=3\Delta/7$ and $eV_2=4\Delta/7$ [panels (c)-(f)])
are not MAR matching voltages of individual
 S$_{\scriptscriptstyle{\mathrm L}}$N$_1$S$_{\scriptscriptstyle{\mathrm C}}$
or S$_{\scriptscriptstyle{\mathrm C}}$N$_2$S$_{\scriptscriptstyle{\mathrm R}}$ parts.
}
\label{fig:fig1}
\end{center}
\end{figure}

An array of alternating superconductor (S) - normal metal (N) islands is a fundamental laboratory
representing a wealth of physical systems ranging from Josephson junction networks
and layered high temperature superconductors to disordered superconducting
films in the vicinity of the superconductor-insulator transition.
Electronic transport in these systems is mediated by Andreev conversion
of a supercurrent into a current of quasiparticles and vice versa at
interfaces between the superconducting and normal regions~\cite{Andreev}.
A fascinating phenomenon benchmarking this mechanism is the enhancement
of the conductivity observed in a single SNS junction at matching voltages constituting
an integer ($m$) fraction of the superconducting gap, $V=2\Delta/(em)$~\cite{SNSexp1,
SNSexp2,SNSexp3,SNSexp4,SNSexp5,SNSexp6,SNSexp7,SNSexp8,SNSexp9,SNSexp10}
due to the effect of multiple Andreev reflection (MAR)~\cite{MAR1,MAR2,MAR3}.
The current-voltage characteristics of diffusive
SNS junctions were discussed in great detail in Refs.~\cite{Shumeiko,Cuevas}.
Further developments were obliged to studies of large arrays comprised of many
SNS junctions~\cite{2DSNS1,2DSNS2,1DSNS,Fritz,TiNperf}.
Experimental results, especially those obtained on the
multiconnected arrays~\cite{2DSNS1,2DSNS2,TiNperf}, indicated clearly that singularities in
transport characteristics cannot be explained by MARs at individual SNS junctions only
and that there is evidently a certain coherence of the Andreev processes that occur at
different NS interfaces.
These findings call for a comprehensive theory of transport in large SNS arrays.

In this article we develop a nonequilibrium theory of electronic transport
in a  series of two diffusive SNS junctions, i.e. an SNSNS junction
and derive the corresponding current-voltage characteristics.
We demonstrate that splitting the normal part of the  SNS junction into two normal islands
that have, in general, different resistances and are
coupled via a small superconducting granule, S$_{\scriptscriptstyle{\mathrm C}}$
leads to the nontrivial physics and emergence of
a new distinct resonant mechanism for the current transfer:
{\it the Synchronized Andreev Transmission} (SAT).
The main component of our consideration is a \textit{nonequilibrium circuit theory}
of the charge transfer across S$_{\scriptscriptstyle{\mathrm C}}$. [The symmetric case with the equal resistances of the normal parts was discussed in detail in~\cite{SNSNS}.  Unfortunately the technique developed there does not allow straightforward generalization onto a nonsymmetric case.]

In the SAT regime the processes of
Andreev conversion at the boundaries of the central superconducting island are correlated:
as a quasiparticle with the energy $\varepsilon$ hits one
NS$_{\scriptscriptstyle{\mathrm C}}$
interface, a quasiparticle with the same energy emerges from the other
S$_{\scriptscriptstyle{\mathrm C}}$N interface and enters
the bulk of the normal island (and vice versa, see Fig.\,1).
This energy synchronization is achieved via over-the-gap
Andreev processes~\cite{1DSNS}, which couple
MARs occurring within the each of the normal islands and
make the quasiparticle distribution at the central island essentially nonequilibrium.
Effectiveness of the synchronization is controlled by the
value of the energy relaxation lengths of both, the quasiparticles crossing
S$_{\scriptscriptstyle{\mathrm C}}$ with energies above $\Delta$,
and the quasiparticles experiencing MAR in the normal parts.
The SAT processes result in spikes in the differential conductivity of the
SNSNS circuit, which appear at resonant values of the \textit{total} applied voltage
$V_{\mathrm{tot}}$  defined by the condition
   \begin{gather}\label{eq:V_position}
       V_{\mathrm{tot}}= \frac{2\Delta}{en}
   \end{gather}
with integer $n$, irrespectively of the details of the distribution of the partial voltages
at the two normal islands.

The article is organized as follows.
In the Section 1.2 we define the system, a diffusive SNSNS junction
which will be a subject of our study.
Section 1.3 is devoted to introduction and description of the employed theoretical tools:
the electronic transport of the system in the resistive state is given by
the Larkin-Ovchinnikov equation in a form of matrix equations
for the Green's functions taken in Keldysh representation.
In Sections 1.4-1.8 we construct an equivalent circuit theory for an SNSNS junction
resulting in the recurrent relations for the spectral current flow in the energy space.
In Section 1.9 we present the original numerical
method enabling us to solve the recurrent relations for the spectral current and obtain
the $I$-$V$ characteristics for the SNSNS junction.
The obtained results are discussed in Section 1.10, where we demonstrate, in particular,
that the SAT-induced features  become dominant in large arrays consisting of
many SNS junctions.

\section{The system}
We consider charge transfer across an
S$_{\scriptscriptstyle{\mathrm L}}$N$_1$S$_{\scriptscriptstyle{\mathrm
C}}$N$_2$S$_{\scriptscriptstyle{\mathrm R}}$
junction, where S$_{\scriptscriptstyle{\mathrm L}}$, S$_{\scriptscriptstyle{\mathrm C}}$,
and S$_{\scriptscriptstyle{\mathrm R}}$
are mesoscopic superconductors with the identical gap $\Delta$; the `edge' superconducting
granules S$_{\scriptscriptstyle{\mathrm L}}$ and S$_{\scriptscriptstyle{\mathrm R}}$
play the role of electrodes, and S$_{\scriptscriptstyle{\mathrm C}}$ is the central
island separating the two normal parts with, in general, different normal resistances.
We discuss the common experimental situation of a diffusive regime where the most of
the energy scales are smaller than $\hbar/\tau$, where $\tau$ is the impurity scattering time.
We assume the size, $L_{\scriptscriptstyle{\mathrm C}}$, of the central island
to be much larger than the
superconducting coherence length $\xi$, hence processes of subgap elastic
cotunneling and/or direct Andreev tunneling~\cite{Deutcher} do not contribute
much to the charge transfer.
In general, this condition ensures that $L_{\scriptscriptstyle{\mathrm C}}$
is large enough so that charges do not accumulate in the central island and
Coulomb blockade effects are irrelevant for the quasiparticle transport.
At the same time $L_{\scriptscriptstyle{\mathrm C}}$ is assumed to be less than the
charge imbalance length, such that we can neglect the coordinate dependence of the
quasiparticle distribution functions across the island S$_{\scriptscriptstyle{\mathrm C}}$.
Additionally, the condition $\ell_{\varepsilon}\gg L_{\scriptscriptstyle{\mathrm C}}$,
where $\ell_{\varepsilon}$ is the energy relaxation length,
implies that quasiparticles with energies $\varepsilon>\Delta$ traverse the
central superconducting island S$_{\scriptscriptstyle{\mathrm C}}$
without any noticeable loss of energy.
The normal parts N$_1$ and N$_2$ are the diffusive normal metals of length
$L_{1,2}>\xi$, and
$L_{1,2}>L_{_\mathrm T}$,
$L_{_\mathrm T}=\sqrt{\hbar D_{_\mathrm N}/\varepsilon}$,
where $D_{_\mathrm N}$ is the diffusion coefficient in the normal metal.
We assume the Thouless energy, $E_{\mathrm {Th}}=\hbar D_{_\mathrm N}/L_{1,2}^2$,
 to be small, $E_{\mathrm {Th}}\ll\Delta$,
and not to exceed the characteristic voltage drops,
$E_{\mathrm {Th}}<eV_{1,2}$.
If these conditions that define the so called incoherent regime~\cite{Shumeiko} are met,
the Josephson coupling between the superconducting islands is suppressed.
And, finally, we let the energy relaxation length in the normal parts N$_1$ and N$_2$
be much larger than their sizes, so that quasiparticles
may experience many incoherent Andreev reflections inside the normal regions.

\section{Theoretical formalism} \label{Sec:Forma}

The current transfer across the SNSNS junction is described by
quasiclassical Larkin-Ovchinnikov (LO) equations for the dirty limit~\cite{Larkin_Ovchinnikov,Kupriyanov-Lukichev}:
\begin{gather}\label{eq:LO}
 -i[\check{H}_{\rm eff}\circ,{\bf \check G}]=\nabla\mathbf{\check{J}},
\quad \check {\mathbf {J}} \cdot \mathbf {n}=\frac{1}{2\sigma_{{_\mathrm S}}R}
[\check{G}_{_\mathrm S}\,,\check{G}_{_\mathrm N}]\,,
\end{gather}
where $\check{H}_{\rm eff}=\check 1 (i\hat\sigma_{\mathrm z}\partial_t-\varphi\hat\sigma_0
+\hat{\Delta})$,
$\mathbf{\check{J}}=D{\bf \check G}\circ\nabla {\bf \check G}$
is the matrix current, the subscripts ``S" and ``N"
denote the superconducting and normal materials, respectively,
``$\circ$'' stands for the time-convolution, $\hat\sigma_{\mathrm i}$ (i\,=\,x,y,z)
are the Pauli matrices,
operating in the Nambu space of $2\times 2$ matrices denoted by `hats',
$\hat\Delta=i\hat\sigma_{\mathrm x}\Im\Delta+i\hat\sigma_{\mathrm y}\Re\Delta$,
and $R$ is the resistance of an NS interface.
The diffusion coefficient $D$ assumes the value
$D_{_\mathrm N}$ in the normal metal and the value
$D_{_\mathrm S}$ in the superconductor, and $\varphi$
is the electrical potential which we calculate self-consistently.
The unit vector $\mathbf n$ is normal to the NS interface and is assumed
to be directed from N to S.
The momentum averaged Green's functions ${\bf \check G}({\bf r}, t, t^{\prime})$
are $2\times 2$ supermatrices in a Keldysh space.
Each element of the Keldysh matrix, labelled with a hat sign, is, in its turn,
a $2\times 2$ matrix in the electron-hole space:
       \begin{equation}
              \label{keldysh-space}
              {\bf \check G} = \left( \begin{array}{cc}
              \hat G^{\mathrm R} & \hat G^{\mathrm K} \\
               0     & \hat G^{\mathrm A}
             \end{array} \right); \hspace{3mm}
             \hat G^{{\mathrm R}({\mathrm A})} = \left( \begin{array}{cc}
              {\cal G}^{{\mathrm R}({\mathrm A})}
              & {\cal F}^{{\mathrm R}({\mathrm A})} \\
             \tilde {\cal F}^{{\mathrm R}({\mathrm A})}
             & \tilde {\cal G}^{{\mathrm R}({\mathrm A})}
          \end{array} \right)\,,
      \end{equation}
${\bf r}$ is the spatial position, $t$ and $t^{\prime}$ are the two time arguments.
The Keldysh component of the Green's function is parametrized as~\cite{Larkin_Ovchinnikov}:
$\hat G^{\mathrm K}=\hat G^{\mathrm R}\circ \hat{f}-\hat{f}\circ \hat G^{\mathrm A}$,
where $\hat f$ is the distribution function matrix, diagonal in Nambu space,
$\hat f\equiv \mathrm{ diag}\,[1-2n_{\mathrm e},1-2n_{\mathrm h}]$,
$n_{\mathrm {e(h)}}$ is the electron (hole) distribution function.
In equilibrium $n_{\mathrm {e(h)}}$ becomes the Fermi function.
And, finally, the Green's function satisfies the normalization
condition ${\bf \check G}^2=\check 1$.

The edge conditions closing Eqs.\,\eqref{eq:LO} are given by the expressions for the
Green's functions in the bulk of the left (L) and right (R) superconducting
leads: $${\bf \check G}_{{\mathrm L}({\mathrm R})}(t,t^{\prime}) =
e^{-i \mu_{{\mathrm L}({{\mathrm R})}}t \hat \tau_3/\hbar}
{\bf \check G}_0(t-t^{\prime}) e^{i \mu_{{\mathrm L}({{\mathrm R})}}t^\prime
\hat\tau_3/ \hbar}\,,$$
the chemical potentials are $\mu_{_\mathrm L}=0$ and $\mu_{_\mathrm R}=eV$.
Here, ${\bf \check G}_0(t)$ is the equilibrium bulk BCS Green's function.

The current density is expressed through the Keldysh component of $\mathbf{\check{J}}$ as
\begin{gather}\label{eq:IVnew}
\mathcal I(t,\mathbf r)=\frac {\pi\sigma_{_{\mathrm N}}}{4}\Tr \hat\sigma_{\mathrm z}
\hat{J}^{\mathrm K}(t,t;\mathbf{r})=\frac {1}{2}\int d\varepsilon
\left[I_{\mathrm e}(\varepsilon)+I_{\mathrm h}(\varepsilon)\right]\,,
\end{gather}
where the spectral currents $I_{\mathrm e}$ and $I_{\mathrm h}$ representing the electron
and hole quasiparticle currents, respectively, are the time Wigner-transforms
of top- and bottom diagonal elements of the  matrix current
$\mathbf{\check J}^{({\mathrm K})}$.

On the normal side of the superconductor-normal metal interface, the Keldysh component
of Eqs.\,\eqref{eq:LO} yield the conservation conditions:
\begin{eqnarray}\label{eqI}
&&\nabla I_{\mathrm {e(h)}}=0,
\\
\label{eq:Ie_def}
&&I_{\mathrm e}(\varepsilon)=\sigma_{\scriptscriptstyle{\mathrm N}}
\left\{D_{\mathrm p}(\varepsilon+u)\nabla\,n_{\mathrm e}(\varepsilon)
-D_{\mathrm m}(\varepsilon+u)\nabla n_{\mathrm h}(\varepsilon+2u)\right\},
\\\label{eq:Ih_def}
&&I_{\mathrm h}(\varepsilon)=\sigma_{\scriptscriptstyle{\mathrm N}}
\left\{D_{\mathrm p}(\varepsilon-u)\,\nabla n_{\mathrm h}(\varepsilon)
-D_{\mathrm m}(\varepsilon-u)\nabla n_{\mathrm e}(\varepsilon-2u)\right\},
\end{eqnarray}
where $u$ is the electrical potential of the adjacent superconductor,
$D_{{\mathrm p}({\mathrm m})}=(D_-\pm D_+)/2$,
  \begin{equation}
    \begin{split}
    &D_+(\varepsilon)=\frac14\Tr[\hat 1-\hat G^{\scriptscriptstyle{\mathrm R}}
    (\varepsilon)\,\hat G^{\scriptscriptstyle{\mathrm A}}(\varepsilon)]\,,
    \\
    &D_-(\varepsilon)=\frac14\Tr[\hat 1-\sigma^z\hat G^{\scriptscriptstyle{\mathrm R}}
(\varepsilon)\,\sigma^z\, \hat G^{\scriptscriptstyle{\mathrm A}}(\varepsilon)]\,,
    \end{split}
 \end{equation}
 and the trace is taken over components in the Nambu-space.
In the bulk of a normal metal,
$\hat G^{\scriptscriptstyle{\mathrm R(A)}}(\varepsilon)\to\pm\hat\sigma_{\mathrm z}$
and $D_+\approx D_-\approx 1$,
so $I_{\mathrm e}=\sigma_{\scriptscriptstyle{\mathrm N}}\nabla n_{\mathrm e}$
and $I_{\mathrm h}=\sigma_{\scriptscriptstyle{\mathrm N}}\nabla n_{\mathrm h}$.

\section{Circuit representation of the boundary conditions}
\begin{figure}[t]
  \center
  \includegraphics[width=\textwidth]{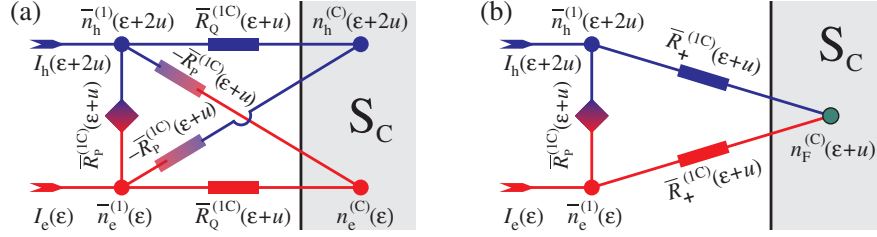}\\
  \caption{Effective circuit for the boundary between the normal metal and the superconductor.
Kirchhoff laws where the role of the potential in the nodes is taken by the electron-
and hole distribution functions give the boundary conditions for the LO equations.
(a) A general nonequilibrium case.
(b) Equivalent circuit for an equilibrium case where quasiparticle distribution functions
in the superconductor are the Fermi-functions, $n_{\mathrm F}$, then
$n_{\mathrm e}(\varepsilon)=n_{\mathrm F}(\varepsilon+u)=n_{\mathrm h}(\varepsilon+2u)$,
used in~\cite{Shumeiko}.
The superconductor electrical potential at the boundary is equal to $u$.
}\label{fig:boundary}
\end{figure}
We start the construction of the circuit theory with the corresponding formulation
of the boundary conditions  for the distribution functions
at the interface between the normal parts and the
central superconducting island.
We consider a stationary situation where the applied voltage
does not depend on time.
Then the Green's functions can be parameterized
near an NS interface as follows:
\begin{equation}\label{eq:sin-cos_representation}
\begin{split}
&[\hat G^{\scriptscriptstyle{\mathrm R}}]_{\mathrm j}(\varepsilon,\varepsilon')
=\hat\sigma_{\mathrm z}\delta_{\varepsilon-\varepsilon'}\cosh\theta_{\mathrm j}(\varepsilon)+
\\
&\hat\sigma_+\,\delta_{\varepsilon-\varepsilon'+2 {\mathrm u}}\sinh\theta_{\mathrm j}(\varepsilon)
-\hat\sigma_-\,\delta_{\varepsilon-\varepsilon'-2u}\sinh\theta_{\mathrm j}(\varepsilon)\,,\\
&G^{\scriptscriptstyle{\mathrm A}}=-\hat\sigma_{\mathrm z}
(G^{\scriptscriptstyle{\mathrm R}})^\dag\hat\sigma_{\mathrm z}\,,
\end{split}
\end{equation}
where $\hat\sigma_\pm=\hat\sigma_{\mathrm x}\pm i\hat\sigma_{\mathrm y}$,
j\,=\,S, N, and $u$ is the electrochemical potential.
The effective diffusion coefficients are correspondingly
$D_+=\cos^2\Imag\theta$ and $D_-=\cosh^2\Real\theta$.
When deriving Eq.\eqref{eq:sin-cos_representation}, we have used the condition
that the Josephson coupling between the superconducting
 islands in the junction is suppressed.
The proximity effect results in an additional term in  Eq.\eqref{eq:sin-cos_representation}
proportional to $\delta(\varepsilon-\varepsilon'-2(u'-u))$, where $u'$
is the potential of the adjacent superconductor involved.

Taking the Keldysh component of the boundary term in Eq.\,\eqref{eq:LO}
we derive the boundary conditions for the currents $I_{\mathrm {e(h)}}$ at the NS interface,
which assume the form of Kirchhoff's laws for the circuit shown in Fig.\,\ref{fig:boundary}(a).
The electron and hole distribution functions take the role of voltages at the nodes.
The equation for an electronic spectral current flowing into the lower left corner node:

\begin{multline}\label{eq:Ie}
I_{\mathrm e}(\varepsilon)=\frac{n_{\mathrm e}^{\scriptscriptstyle{\scriptscriptstyle{\mathrm {(C)}}}}(\varepsilon)
-n_{\mathrm e}^{\scriptscriptstyle{\mathrm {(1)}}}(\varepsilon)}{R_{\scriptscriptstyle{\rm Q}}^{\scriptscriptstyle{\scriptscriptstyle{\mathrm {(1C)}}}}(\varepsilon+u)}+
\frac{n_{\mathrm h}^{\scriptscriptstyle{\scriptscriptstyle{\mathrm {(C)}}}}(\varepsilon+2u)- n_{\mathrm e}^{\scriptscriptstyle{\mathrm {(1)}}}(\varepsilon)}
{[-R_{\scriptscriptstyle{\mathrm P}}^{\scriptscriptstyle{\scriptscriptstyle{\mathrm {(1C)}}}}(\varepsilon+u)]}+
\\
\frac{n_{\mathrm h}^{\scriptscriptstyle{\mathrm {(1)}}}(\varepsilon+2u)- n_{\mathrm e}^{\scriptscriptstyle{\mathrm {(1)}}}(\varepsilon)}
{R_{\scriptscriptstyle{\mathrm P}}^{\scriptscriptstyle{\scriptscriptstyle{\mathrm {(1C)}}}}(\varepsilon+u)}\,.
\end{multline}
The equation for the hole current going into the top left node of the circuit of
the Fig.\,\ref{fig:boundary}(a)
is easily obtained analogously to \eqref{eq:Ie}
with the aid of the additional transformation $\varepsilon\to \varepsilon-2u$ i.e.
by shifting all the energies over $-2u$:

\begin{multline}\label{eq:Ih}
I_{\mathrm h}(\varepsilon)=\frac{n_{\mathrm h}^{\scriptscriptstyle{\scriptscriptstyle{\mathrm {(C)}}}}(\varepsilon)
-n_{\mathrm h}^{\scriptscriptstyle{\mathrm {(1)}}}(\varepsilon)}{R_{\scriptscriptstyle{\mathrm Q}}^{\scriptscriptstyle{\scriptscriptstyle{\mathrm {(1C)}}}}(\varepsilon-u)}+
\frac{n_{\mathrm e}^{\scriptscriptstyle{\scriptscriptstyle{\mathrm {(C)}}}}(\varepsilon-2u)- n_{\mathrm h}^{\scriptscriptstyle{\mathrm {(1)}}}(\varepsilon)}
{[-R_{\scriptscriptstyle{\mathrm P}}^{\scriptscriptstyle{\scriptscriptstyle{\mathrm {(1C)}}}}(\varepsilon-u)]}+
\\
\frac{n_{\mathrm e}^{\scriptscriptstyle{\mathrm {(1)}}}(\varepsilon-2u)- n_{\mathrm h}^{\scriptscriptstyle{\mathrm {(1)}}}
(\varepsilon)}{R_{\scriptscriptstyle{\mathrm P}}^{\scriptscriptstyle{\scriptscriptstyle{\mathrm {(1C)}}}}(\varepsilon-u)}\,.
\end{multline}

In an equilibrium the quasiparticles in the superconductor follow the Fermi
distribution,
then $n_{\mathrm e}^{\scriptscriptstyle{\scriptscriptstyle{\mathrm {(C)}}}}(\varepsilon)=n_{\scriptscriptstyle{\mathrm F}}(\varepsilon+u)
=n_{\mathrm h}^{\scriptscriptstyle{\scriptscriptstyle{\mathrm {(C)}}}}(\varepsilon+2u)$
and Eqs.\eqref{eq:Ie}-\eqref{eq:Ih} reduce to
\begin{gather}\label{eq:Ie_eq}
I_{\mathrm e}(\varepsilon)=\frac{n_{\scriptscriptstyle{\mathrm F}}(\varepsilon-u)-n_{\mathrm e}^{\scriptscriptstyle{\mathrm {(1)}}}
(\varepsilon)}{R_+^{\scriptscriptstyle{\scriptscriptstyle{\mathrm {(1C)}}}}(\varepsilon+u)}+
\frac{n_{\mathrm h}^{\scriptscriptstyle{\mathrm {(1)}}}(\varepsilon+2u)- n_{\mathrm e}^{\scriptscriptstyle{\mathrm {(1)}}}
(\varepsilon)}{R_{\scriptscriptstyle{\mathrm P}}^{\scriptscriptstyle{\scriptscriptstyle{\mathrm {(1C)}}}}(\varepsilon+u)}\,,
\end{gather}
where the interjacent resistances are defined as ${R}_{{_\mathrm Q({_\mathrm P})}}^{-1}(\varepsilon)=\{{R}^{-1}_{-}(\varepsilon)\pm \bar{R}^{-1}_{+}(\varepsilon)\}/2$; here $1/R_\pm(\varepsilon)=[N_2\,N_1\mp M_2^\pm\,M_1^\pm]/R$, $N_{j}(\varepsilon)=\Real\cosh\theta_{j}$, $M_j^+(\varepsilon)+i\,M_j^-(\varepsilon)=\sin\theta_j$ and $j=1,2$ labels the different sides of the interface.


The circuit representation of Eq.\eqref{eq:Ie_eq} is shown
in Fig.\,\ref{fig:boundary}(b).
This is the boundary conditions and the corresponding circuit used in Ref.\cite{Shumeiko}.

\section{Conductance renormalization procedure}
\begin{figure}[t]
  \center
  \includegraphics[width=90mm]{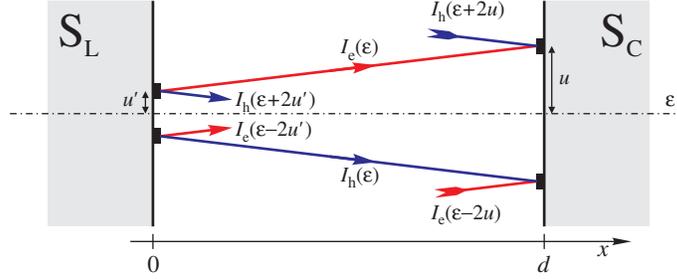}
  \\
  \caption{Illustration of the spectral currents flow in the normal metal
[white area] surrounded by the superconductors [grey area].
The black boxes at the interfaces encode the boundary conditions picture
like it is shown in Fig.\ref{fig:boundary}.} \label{fig:renormalization}
\end{figure}

We consider the normal metal between the left superconducting
lead S$_{\scriptscriptstyle{\mathrm L}}$
and the superconducting island, S$_{\scriptscriptstyle{\mathrm C}}$,
see Fig.\,\ref{fig:renormalization}.
The boundary conditions, Eqs.\eqref{eq:Ie}-\eqref{eq:Ih},
relate electron and hole distribution functions at the right NS and left NS interfaces.
Below we relate electron and hole distribution functions at $x=0$ and $x=d$
building the effective circuit, where $d$ is the length of the normal layer.
At the first step we neglect the proximity effect change of the junction resistance
and take $D_\pm=1$ everywhere in the normal layer.
Then, $I_{\mathrm {e(h)}}=\sigma_{1}\,\nabla n_{\mathrm {e(h)}}$,
$\triangle n_{\mathrm {e(h)}}=0$ and therefore
\begin{gather}\label{eq:R_NN}
I_{\mathrm {e(h)}}(\varepsilon)=[n_{\mathrm {e(h)}}(d,\varepsilon)
-n_{\mathrm {e(h)}}(x=0,\varepsilon)]/R_1\,,
\end{gather}
where $R_{1}=\sigma_{1}/d$ is the normal resistance of the N$_1$-layer.
Eq.\eqref{eq:R_NN} resembles the Ohm law for the resistor $R_1$,
but where the role of the voltages play the distribution functions
at the ends of the resistor.  Eq.\eqref{eq:R_NN}
is approximate because it neglects the proximity renormalization
of the normal layer conductivity \cite{Volkov}.
It was shown in Ref.\cite{Shumeiko} for a SNS junction that the
replacement of $n_{\mathrm {e(h)}}(x=\{0,d\})$  by the properly chosen proximity
renormalized distribution functions makes Eq.\eqref{eq:R_NN} accurate.
We show below that this idea is applicable when electron and hole distribution
functions in the superconductors essentially deviate from the Fermi
functions and when the electron and hole currents can not be in general
related by a shift of the energy like in SNS junction.

At the left NS-interface the spectral currents,
$I_{\mathrm e}(\varepsilon)$ and $I_{\mathrm h}(\varepsilon+2u)$
are related by the Andreev process, see Fig.\ref{fig:boundary}a.
It follows from Eqs.\eqref{eqI}-\eqref{eq:Ih_def} that the combination
of the quasiparticle currents,
$I_{\pm}^{\scriptscriptstyle{\mathrm {(1C)}}}(\varepsilon)=I_{\mathrm e}(\varepsilon)\pm I_{\mathrm h}(\varepsilon+2u)
=\sigma_{1}D_\pm^{\scriptscriptstyle{\mathrm {(1C)}}}
(\varepsilon+u)\nabla n^{\scriptscriptstyle{\mathrm {(1C)}}}_\pm(\varepsilon+u)$,
conserve in the normal metal:
$\nabla I_{\pm}^{\scriptscriptstyle{\mathrm {(1C)}}}=0$.
Integrating the last equation over $x$ we get,
\begin{gather}\label{eq:I_pm_xd}
I_\pm^{\scriptscriptstyle{\mathrm {(1C)}}}\int_0^d \frac{dx}{D_{\pm}^{\scriptscriptstyle{\mathrm {(1C)}}}(x)}=\sigma_{1}\,[n_\pm(d)-n_\pm(x)],
\end{gather}
where $n_\pm^{\scriptscriptstyle{\mathrm {(1C)}}}(\varepsilon)=n_{\mathrm e}(\varepsilon)\pm n_{\mathrm h}(\varepsilon+2u)$.
Eq.\eqref{eq:I_pm_xd} can be equivalently rewritten:
\begin{gather}\label{eq:I_pm_as}
I_\pm^{\scriptscriptstyle{\mathrm {(1C)}}} (d-x)=\sigma_{1}\,[\bar n^{\scriptscriptstyle{\mathrm {(1C)}}}_\pm(d)-n^{\scriptscriptstyle{\mathrm {(1C)}}}_\pm(x)],
\\\label{eq:bar_n_pm_1}
\bar n^{\scriptscriptstyle{\mathrm {(1C)}}}_\pm(d)\equiv n^{\scriptscriptstyle{\mathrm {(1C)}}}_\pm(d)-m_\pm^{\scriptscriptstyle{\mathrm {(1C)}}}I_\pm^{\scriptscriptstyle{\mathrm {(1C)}}}(\varepsilon+u),
\end{gather}
where
\begin{gather}\label{eq:m_pm_def}
m_\pm^{\scriptscriptstyle{\mathrm {(1C)}}}=\frac{1}{\sigma_{1}}\int_{0}^{d} \left(\frac{1}{D_{\pm}^{\scriptscriptstyle{\mathrm {(1C)}}}(x)}-1\right)dx.
\end{gather}
Here the variable $x$ occupies the domain $\xi_{\scriptscriptstyle{\mathrm N}}\ll x\ll d-\xi_{\scriptscriptstyle{\mathrm N}}$
where the Cooper pair wave functions from the left and right superconductors,
see Fig.\ref{fig:renormalization}, do not overlap. At these values of $x$,
the angle $\theta(x)\to 0$,  $D_{\pm}(x)\to 1$ and  we can therefore substitute
$x$ by $0$ in the integral written in Eq.\eqref{eq:I_pm_xd}.

Taking into account that $I_\pm^{\scriptscriptstyle{\mathrm {(1C)}}}(\varepsilon+u)=I_{\mathrm e}(\varepsilon)\pm I_{\mathrm h}(\varepsilon+2u)$
we finally get the following important result:
\begin{gather}\label{eq:Iex1}
    (d-x)\,I_{\mathrm e}(\varepsilon)=\sigma_{1}[\bar n^{\scriptscriptstyle{\mathrm {(1C)}}}_{\mathrm e}(d)-n^{\scriptscriptstyle{\mathrm {(1C)}}}_{\mathrm e}(x)],
\end{gather}
where
\begin{gather}\label{eq:bar_ne}
\bar n^{\scriptscriptstyle{\mathrm {(1C)}}}_{\mathrm e}(d)=n^{\scriptscriptstyle{\mathrm {(1C)}}}_{\mathrm e}(d)-I_{\mathrm e}(\varepsilon)\,m_{\mathrm e}^{\scriptscriptstyle{\mathrm {(1C)}}}(\varepsilon+u)-I_{\mathrm h}
(\varepsilon+2u)\,m_{\mathrm h}^{\scriptscriptstyle{\mathrm {(1C)}}}(\varepsilon+u)\,.
\end{gather}
Here $m_{\mathrm {e(h)}}^{\scriptscriptstyle{\mathrm {(1C)}}}=[m_+^{\scriptscriptstyle{\mathrm {(1C)}}}\pm m_-^{\scriptscriptstyle{\mathrm {(1C)}}}]/2$.

Applying the procedure, Eqs.\eqref{eq:I_pm_xd}--\eqref{eq:bar_ne},
to the left NS interface in Fig.\ref{fig:renormalization}, we find for $\xi_{\scriptscriptstyle{\mathrm N}}\ll x\ll d-\xi_{\scriptscriptstyle{\mathrm N}}$:
\begin{gather}\label{eq:Iex2}
    x\,I_{\mathrm e}(\varepsilon)=\sigma_{1}[n^{\scriptscriptstyle{\mathrm {(1L)}}}_{\mathrm e}(x)
-\bar n^{\scriptscriptstyle{\mathrm {(1L)}}}_{\mathrm e}(x=0)]\,,
\end{gather}
where
\begin{multline}\label{eq:bar_nee}
\bar n^{\scriptscriptstyle{\mathrm {(1L)}}}_{\mathrm e}(x=0)
=n^{\scriptscriptstyle{\mathrm {(1L)}}}_{\mathrm e}(x=0)+I_{\mathrm e}(\varepsilon)\,m_{\mathrm e}^{\scriptscriptstyle{\mathrm {(1L)}}}(\varepsilon+u')+
\\
I_{\mathrm h}(\varepsilon+2u')\,m_{\mathrm h}^{\scriptscriptstyle{\mathrm {(1L)}}}(\varepsilon+u').
\end{multline}
Here $m_{\mathrm {e(h)}}^{\scriptscriptstyle{\mathrm {(1L)}}}=[m_+^{\scriptscriptstyle{\mathrm {(1L)}}}\pm m_-^{\scriptscriptstyle{\mathrm {(1L)}}}]/2$,
\begin{gather}\label{eq:bar_n_pm_2}
m_\pm^{\scriptscriptstyle{\mathrm {(1L)}}}=\frac{1}{\sigma_{1}}\int_{0}^d \left(\frac{1}{D_{\pm}^{\scriptscriptstyle{\mathrm {(1L)}}}(x)}-1\right)dx.
\end{gather}

Eqs.\eqref{eq:Iex1}-\eqref{eq:Iex2} show how $n^{(1)}_{\mathrm e}$ depends on $x$
in the central part of the normal layer in Fig.\ref{fig:renormalization}.
Eq.\eqref{eq:Iex1} must be consistent with Eq.\eqref{eq:Iex2}.
The only way to satisfy this condition is the following one:
\begin{gather}\label{eq:Ohm_e}
I_{\mathrm e}(\varepsilon)=\frac{\bar n^{\scriptscriptstyle{\mathrm {(1C)}}}_{\mathrm e}(d)-\bar n^{\scriptscriptstyle{\mathrm {(1L)}}}_{\mathrm e}(x=0)}{R_{1}},
\end{gather}
where $R_{1}=d/\sigma_{1}$ is the normal resistance of the N-layer
and we used that $n^{\scriptscriptstyle{\mathrm {(1L)}}}_{\mathrm e}(x)=n^{\scriptscriptstyle{\mathrm {(1C)}}}_{\mathrm e}(x)$.
The condition Eq.\eqref{eq:Ohm_e}
resembles the Ohm law. It allows to relate the distribution functions at $x=0$ and $x=d$.

Similar condition holds for $I_{\mathrm h}$:
     \begin{gather}\label{eq:Ohm_h}
          I_{\mathrm h}(\varepsilon)=\frac{\bar n^{\scriptscriptstyle{\mathrm {(1C)}}}_{\mathrm h}-\bar n^{\scriptscriptstyle{\mathrm {(1L)}}}_{\mathrm h}}{R_{1}},
      \end{gather}
where
\begin{gather}\label{eq:bar_neee}
\bar n^{\scriptscriptstyle{\mathrm {(1C)}}}_{\mathrm h}=n^{\scriptscriptstyle{\mathrm {(1C)}}}_{\mathrm h}(d)
-m_{\mathrm h}^{\scriptscriptstyle{\mathrm {(1C)}}}(\varepsilon-u)\,I_{\mathrm e}(\varepsilon-2u)
-m_{\mathrm e}^{\scriptscriptstyle{\mathrm {(1C)}}}(\varepsilon-u)\,I_{\mathrm h}(\varepsilon)\,,
\\\label{eq:bar_nhhe}
\bar n^{\scriptscriptstyle{\mathrm {(1L)}}}_{\mathrm h}=n^{\scriptscriptstyle{\mathrm {(1L)}}}_{\mathrm h}
+m_{\mathrm h}^{\scriptscriptstyle{\mathrm {(1L)}}}(\varepsilon-u')\,I_{\mathrm e}(\varepsilon-2u')+
m_{\mathrm e}^{\scriptscriptstyle{\mathrm {(1L)}}}(\varepsilon-u')\,I_{\mathrm h}(\varepsilon)]\,.
\end{gather}

The last step is the formulation of the boundary conditions
at the NS interfaces in terms of the distribution functions with bars.
Using the $I_{\pm}$ notations we can rewrite the boundary conditions,
Eqs.\eqref{eq:Ie}-\eqref{eq:Ih}, in the compact form
\begin{gather}\label{eq:boundary_PM}
I_\pm^{\scriptscriptstyle{\mathrm {(1C)}}}(\varepsilon)
=\frac{n^{\scriptscriptstyle{\mathrm {(C)}}}_\pm(\varepsilon)-n^{\scriptscriptstyle{\mathrm {(1C)}}}_\pm(\varepsilon)}{R_\pm^{\scriptscriptstyle{\mathrm {(1C)}}}(\varepsilon+u)}.
\end{gather}
Then it follows from Eq.\eqref{eq:I_pm_as} that we can write:
\begin{gather}\label{eq:boundary_PM_bar}
I_\pm^{\scriptscriptstyle{\mathrm {(1C)}}}(\varepsilon+u)
=\frac{n^{\scriptscriptstyle{\mathrm {(C)}}}_\pm-\bar n^{\scriptscriptstyle{\mathrm {(1C)}}}_\pm}{\bar R_\pm^{\scriptscriptstyle{\mathrm {(1C)}}}(\varepsilon+u)},
\\
\bar R_\pm^{\scriptscriptstyle{\mathrm {(1C)}}}(\varepsilon)={m_\pm^{\scriptscriptstyle{\mathrm {(1C)}}}(\varepsilon)+R_\pm^{\scriptscriptstyle{\mathrm {(1C)}}}(\varepsilon)}.
\end{gather}
The same form has the boundary condition at $x=0$:
\begin{gather}
I_\pm^{\scriptscriptstyle{\mathrm {(1L)}}}(\varepsilon+u')=\frac{\bar n^{\scriptscriptstyle{\mathrm {(1L)}}}_\pm-n^{\scriptscriptstyle{({\mathrm L})}}_\pm}{\bar R_\pm^{\scriptscriptstyle{\mathrm {(1L)}}}(\varepsilon+u')},
\\
\bar R_\pm^{\scriptscriptstyle{\mathrm {(1L)}}}(\varepsilon)={m_\pm^{\scriptscriptstyle{\mathrm {(1L)}}}(\varepsilon)+R_\pm^{\scriptscriptstyle{\mathrm {(1L)}}}(\varepsilon)}.
\end{gather}
It follows that the physical meaning of  $m_\pm$ terms is the proximity
effect contribution to the NS interface resistance, see \cite{Shumeiko}.

It is more convenient to work with the boundary conditions for $I_{\mathrm {e(h)}}$
rather then with those for $I_\pm$.  Then one can use Eqs.\eqref{eq:Ie},\eqref{eq:Ih}
but with $n^{\scriptscriptstyle{\mathrm {(1L)}}}_{\mathrm {e(h)}}\to \bar n^{\scriptscriptstyle{\mathrm {(1L)}}}_{\mathrm {e(h)}}$
and $R_{\mathrm {Q(P)}}^{\scriptscriptstyle{\mathrm {(1C)}}}\to \bar R_{\mathrm {Q(P)}}^{\scriptscriptstyle{\mathrm {(1C)}}}$,
where, for example,
$\bar R_{\mathrm {Q(P)}}^{\scriptscriptstyle{\mathrm {(1C)}}}
=2\bar R_-^{\scriptscriptstyle{\mathrm {(1C)}}}\bar R_+^{\scriptscriptstyle{\mathrm {(1C)}}}/[\bar R_+^{\scriptscriptstyle{\mathrm {(1C)}}}\pm
\bar R_-^{\scriptscriptstyle{\mathrm {(1C)}}}]$.

\section{Retarded and advanced Greens functions evolution in normal metals and superconductors}

Having formulated the boundary conditions for the distribution functions
we turn now to  advanced and retarded Greens functions behaviors.
\begin{figure}[b]
\center
  \includegraphics[width=90mm]{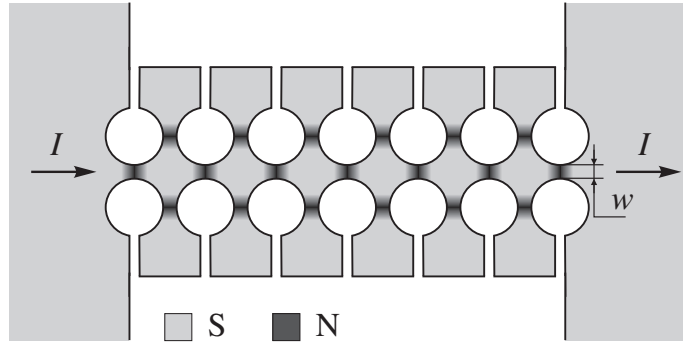}\\
  \caption{Typical experimental array of SNS junctions~\cite{2DSNS1,1DSNS}.
This type of the link enables us to use the rigid boundary conditions
for the retarded and advanced Greens functions.
}\label{fig_SNS}
\end{figure}

Normal layers in experimental SNS junctions and SNS arrays, see Ref.\cite{2DSNS1,1DSNS},
connect with superconductors like it is  shown in Figs.\ref{fig_SNS}.
The junctions of this type are usually referred to as ``weak-links''.~\cite{Likharev,Beenakker}
Boundary conditions for retarded and advanced Greens functions, Eq.\eqref{eq:LO},
can be simplified in this case: retarded and advanced Greens functions
at superconducting sides of NS boundaries can be substituted by Greens
function from the bulk of the superconductors.
These ``rigid'' boundary conditions approximation is reasonable because
the magnitude of the current is much smaller than the critical current of the superconductor
[this is assumed] and the current entering the
superconductor from narrow normal metal wire with the width comparable with the Cooper pair size.
There are  also other cases when the rigid boundary conditions are correct, for example,
if the NS boundary has the small transparency due to, e.g., an insulator layer at the NS interface.

The recipe telling how one should evaluate $\theta(x)$ in the normal metal near the NS
boundary, where the rigid boundary conditions hold, can be taken from
e.g. Ref.\cite{Shumeiko}, and we reproduce briefly their result for the completeness.
We will write down $\theta(x)$ near the right NS boundary (see
Fig.\ref{fig:renormalization} )  taking $u'=0$.
In the superconductor, $\theta_{\scriptscriptstyle{\mathrm S}}=\mathrm{atanh}\,(\Delta/\varepsilon)$,
where $\Delta$ is the gap. The value of $\theta_{\scriptscriptstyle{\mathrm N}}=\theta(x=0)$ in the normal
metal side should be found from the equation:
     \begin{gather}\label{eq:121}
             W\sqrt{\frac{i\Delta}{\varepsilon+i/2\tau_\sigma}}\sinh(\theta_{\scriptscriptstyle{\mathrm N}}-\theta_{\scriptscriptstyle{\mathrm S}})+2\sinh\frac {\theta_{\scriptscriptstyle{\mathrm N}}}
              2=0,
     \end{gather}
where $W=R_{\scriptscriptstyle{\mathrm\Delta}}/R_{\scriptscriptstyle{\mathrm{NS}}}$, $R_{\scriptscriptstyle{\mathrm\Delta}}=
\xi_{\scriptscriptstyle{\mathrm\Delta}}/\sigma_{\scriptscriptstyle{\mathrm N}}$, ($\xi_{\scriptscriptstyle{\mathrm\Delta}}=\sqrt{D/\Delta}$)
is the resistance of the normal metal layer with the width $\xi_{\scriptscriptstyle{\mathrm\Delta}}$. Here $\tau_\sigma$
is the pair breaking rate\cite{Tinkham} [e.g., induced by electron-phonon or electron-electron
interactions] and $R_{\scriptscriptstyle{\mathrm{NS}}}$ is the normal resistance of the interface.
Then the solution for $\theta(x>0)$ is the following:
\begin{gather}\label{eq:122}
    \tanh\frac\theta 4= \exp\left(-\frac{x}{\xi_{\varepsilon}\sqrt i}\right)\tanh\frac{\theta_{\scriptscriptstyle{\mathrm N}}} 4,
\end{gather}
where $\xi_{\varepsilon}=\sqrt{D/[2(\varepsilon+i/2\tau_\sigma)]}$.
The effective conductances $\bar g_\pm$ should be expressed through $\theta$
found from Eqs.\eqref{eq:121}-\eqref{eq:122}.

\section{Spectral current flow through the superconducting grains}
\begin{figure}[t]
\center
  \includegraphics[width=\textwidth]{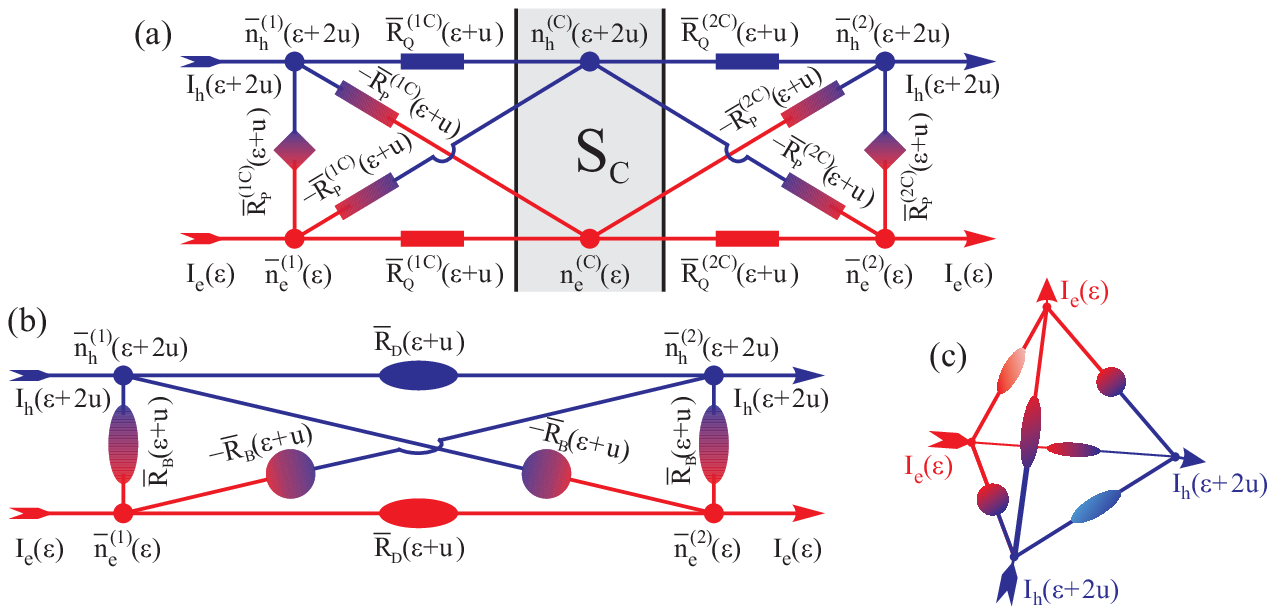}\\
  \caption{(a) Effective circuit representing current conversion at the interfaces of
  the central superconducting island S$_{\scriptscriptstyle{\mathrm C}}$.
Resistors, $R_{\scriptscriptstyle{\mathrm{P}}}$
  and  $R_{\scriptscriptstyle{\mathrm{Q}}}$ stand for an Andreev-
  and a normal processes respectively.
The role of voltages at the nodes is played by the electron and hole
  distribution functions.
(b) An illustration of the boundary conditions
Eqs.\eqref{eq:Ie_boundary}-\eqref{eq:G_DI} in terms of a pyramid-circuit is given in
this figure. Electron and hole currents entering the left side of the
pyramid flow in one normal layer, the right currents flow in the other normal layer.
The effective resistance $\bar R_{\mathrm D}$ describes the ``direct'' quasiparticle
transmission from one normal layer to the other through the superconductor
and the resistance $\bar R_{\mathrm B}$ describes Andreev processes. c) Equivalent 3D-sketch of the circuit (b).}
\label{fig:NSN}
\end{figure}

The circuit shown in Fig.\ref{fig:NSN}a is the graphic representation
of  the boundary conditions to Eq.\eqref{eq:LO} at the edges of the superconducting island.
It is constructed from the circuit units shown in Fig.\ref{fig:boundary}a.
We consider the case where the size of the superconducting island is less than
the charge imbalance length, and therefore the coordinate dependence
of the quasiparticle distribution functions at the island can be neglected.
Solving the Kirchhoff equations for the circuit shown in Fig.\ref{fig:NSN}a
we exclude the quasiparticle distribution functions corresponding
to the superconducting island and express the spectral currents through
the quasiparticle distribution functions in the normal layers:
\begin{gather}\label{eq:Ie_boundary}
I_{\mathrm e}(\varepsilon)=\frac{\bar n_{\mathrm e}^{(2)}(\varepsilon)
-\bar n_{\mathrm e}^{(1)}(\varepsilon)}{\bar R_{\scriptscriptstyle{\mathrm D}}(\varepsilon+u)}+
\frac{\bar n_{\mathrm h}^{(2)}(\varepsilon+2u)
-\bar n_{\mathrm h}^{(1)}(\varepsilon+2u)}{\bar R_{\scriptscriptstyle{\mathrm B}}(\varepsilon+u)},
\\\label{eq:Ih_boundary}
I_{\mathrm h}(\varepsilon)=\frac{\bar n_{\mathrm h}^{(2)}(\varepsilon)
-\bar n_{\mathrm h}^{(1)}(\varepsilon)}{\bar R_{\scriptscriptstyle{\mathrm D}}(\varepsilon-u)}+
\frac{\bar n_{\mathrm e}^{(2)}(\varepsilon-2u)
-\bar n_{\mathrm e}^{(1)}(\varepsilon-2u)}{\bar R_{\scriptscriptstyle{\mathrm B}}(\varepsilon-u)},
\\\label{eq:G_DI}
\bar R_{\scriptscriptstyle{\mathrm {D(B)}}}=2\left[\frac{1}{\bar R_+^{\scriptscriptstyle{\scriptscriptstyle{\mathrm {(1C)}}}}+
\bar R_+^{\scriptscriptstyle{\scriptscriptstyle{\mathrm {(2C)}}}}}\pm\frac{1}{\bar R_-^{\scriptscriptstyle{\scriptscriptstyle{\mathrm {(1C)}}}}+\bar R_-^{\scriptscriptstyle{\scriptscriptstyle{\mathrm {(2C)}}}}} \right]^{-1}\,.
\end{gather}
The effective resistance $\bar R_{\scriptscriptstyle{\mathrm D}}$ describes the ``direct'' quasiparticle transmission
from one normal layer to the other through the superconductor and the resistance
$\bar R_{\scriptscriptstyle{\mathrm B}}$ describes the Andreev processes, see Fig.\ref{fig:NSN}b.
Note that the direct and indirect transmissions here are different from the so-called ``elastic co-tunneling''
and ``crossed Andreev tunneling'' \cite{Deutcher,Chtchelkatchev_spin_filter} processes
where Bogoliubov quasiparticles tunnel below the gap through a thin
(with the width of the order of the Cooper pair size) superconducting layer.
The probability of these tunneling processes decreases exponentially
if the width of the superconducting layer exceeds the Cooper pair size.
and they occur  without generating supercurent across a superconductor
(the supercurrent flows ``virtually'').
The size of superconucting islands of the SNS arrays that we consider here exceed well the Cooper pair size,
and the current of the quasiparticles with the energies below the hap converts at the NS interface
into the supercurrent across the S-islands and then transforms again into
the quasiparticle current at the opposite SN-interface.

\section{Recurrent relations}
\begin{figure}[t]
\begin{center}
\includegraphics[width=\textwidth]{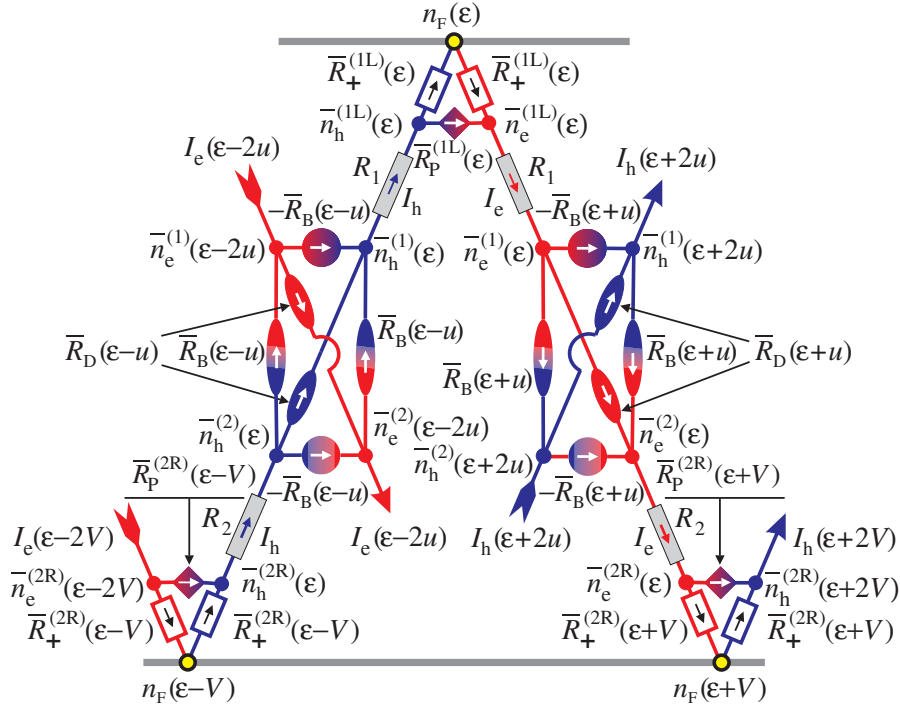}
\caption{MAR in a SNSNS array. The graph shows the effective
circuit for quasiparticle currents $I_{\mathrm e}$ and $I_{\mathrm h}$ in the energy space.
The role of voltages here play quasiparticle distribution functions.
Boxes, triangles and ovals play the role of effective
resistances that come from Usadel equations and their boundary conditions.
\label{fig:SNSNS}}
\end{center}
\end{figure}

We have demonstrated that there is a direct correspondence between the effective electric
circuit and the solution of the Usadel equations with the appropriately chosen boundary conditions.
The effective circuit describing transport in SNSNS-array is shown
in Fig.\ref{fig:SNSNS}. We choose the direction of the current
flow in such a way that the electron, $I_{\mathrm e}$, and the hole, $I_{\mathrm h}$,
currents go in opposite directions.  The expression for the total current then assumes the form:
\begin{gather}\label{eq:IVnew1}
\mathcal I(V)=-\frac {1}{2e}\int d\varepsilon \left(I_{\mathrm e}+I_{\mathrm h}\right).
\end{gather}
The spectral currents $I_{\mathrm e}$ and $I_{\mathrm h}$ satisfy in general the relation: $I_{\mathrm e}(\varepsilon)=-I_{\mathrm h}(\varepsilon)|_{V\to-V}$.
Similarly, $n_{\mathrm e}(\varepsilon)=n_{\mathrm h}(\varepsilon)|_{V\to -V}$, ensuring the identity $\mathcal I(-V)=-\mathcal I(V)$.

Writing down the Kirchofs equations for potential distribution at the circuit
in Fig.\,\ref{fig:SNSNS}
we arrive at the recurrent relations, see Appendix \ref{sec:Appendix}:
\begin{gather}\label{eq:recurrent1}
\begin{split}
\mathcal{R}(\varepsilon,-{\mathrm u},- V)
I_{\mathrm h}(\varepsilon)-&\rho^{(\circ)}({\varepsilon-{\mathrm u}})
I_{\mathrm e}(\varepsilon-2{\mathrm u})-
\\
\rho^{(\triangleright)}(\varepsilon)I_{\mathrm e}(\varepsilon)&
-\rho^{(\triangleleft)}(\varepsilon -V)I_{\mathrm e}(\varepsilon- V)
=n_{\scriptscriptstyle{\mathrm F}}(\varepsilon)-n_{\scriptscriptstyle{\mathrm F}}(\varepsilon-V),
\end{split}
\\ \label{eq:recurrent2}
\begin{split}
\mathcal{R}(\varepsilon,{\mathrm u}, V)I_{\mathrm e}(\varepsilon)&
- \rho^{(\circ)}(\varepsilon+{\mathrm u})I_{\mathrm h}(\varepsilon+2{\mathrm u})-
\\
\rho^{(\triangleright)}(\varepsilon) I_{\mathrm h}(\varepsilon)&
- \rho^{(\triangleleft)}(\varepsilon+ V)I_{\mathrm h}(\varepsilon+2V)
=n_{\scriptscriptstyle{\mathrm F}}(\varepsilon+V)-n_{\scriptscriptstyle{\mathrm F}}(\varepsilon).
\end{split}
\end{gather}
Here the effective resistance
$
\mathcal{R}=R_{{1}}+R_{2}+\rho^{(\triangleright\circ\triangleleft)},
$
where
\begin{gather}
\rho^{(\triangleright\circ\triangleleft)}=
(1/2)\sum_{\alpha=\pm}
\{\bar R_{\alpha,\varepsilon}^{({{\scriptscriptstyle{\mathrm {1L}}}})}+
  \bar R_{\alpha,\varepsilon+{\mathrm u}}^{({{\scriptscriptstyle{\mathrm {1C}}}})}+
  \bar R_{\alpha,\varepsilon+{\mathrm u}}^{({{\scriptscriptstyle{\mathrm {2C}}}})}+
  \bar R_{\alpha,\varepsilon+\scriptscriptstyle{\mathrm V}}^{({{\scriptscriptstyle{\mathrm {2R}}}})}\},
  \\
\rho^{(\circ)}=(1/2)\{
 \bar R_{+}^{({{\scriptscriptstyle{\mathrm {1C}}}})}
+\bar R_{+}^{({{\scriptscriptstyle{\mathrm {2C}}}})}
-\bar R_{-}^{({{\scriptscriptstyle{\mathrm {1C}}}})}
-\bar R_{-}^{({{\scriptscriptstyle{\mathrm {2C}}}})}\},
\\
\rho^{(\triangleleft)}=(1/2)\{
 \bar R_{+}^{({{\scriptscriptstyle{\mathrm {2R}}}})}
-\bar R_{-}^{({{\scriptscriptstyle{\mathrm {2R}}}})}\},
\\
\rho^{(\triangleright)}=(1/2)\{
 \bar R_{+}^{({{\scriptscriptstyle{\mathrm {1L}}}})}
-\bar R_{-}^{({{\scriptscriptstyle{\mathrm {1L}}}})}\}.
\end{gather}
In the normal state of the array (or if $|\varepsilon|\gg\Delta$) $\mathcal{R}$
reduces to a normal resistance of the array whereas $\rho^{(\triangleleft)}$ and
$\rho^{(\triangleright)}$ vanish.
Then we find from Eqs.\eqref{eq:recurrent1}-\eqref{eq:recurrent2}
that $I_{\mathrm h}(\varepsilon)=
[n_{\scriptscriptstyle{\mathrm F}}(\varepsilon)-n_{\scriptscriptstyle{\mathrm F}}(\varepsilon-V)]
/\mathcal{R}$, and $I_{\mathrm e}(\varepsilon)=
[n_{\scriptscriptstyle{\mathrm F}}(\varepsilon+V)-n_{\scriptscriptstyle{\mathrm F}}(\varepsilon)]
/\mathcal{R}$
that with Eq.\eqref{eq:IVnew1}
reproduces the Ohm's law, $\mathcal I=V/\mathcal{R}$.

It is easy to find the island potential in the case of symmetrical array when the transmitivities of the island-normal metal interfaces are equal as well as the transmitivities of the lead-normal metal interfaces and $R_{1}=R_{2}$. Then the resistances $\bar R_{\pm}^{({{\scriptscriptstyle{\mathrm {1L}}}})}=\bar R_{\pm}^{({{\scriptscriptstyle{\mathrm {1R}}}})}$, $\bar R_{\pm}^{({{\scriptscriptstyle{\mathrm {1C}}}})}=\bar R_{\pm}^{({{\scriptscriptstyle{\mathrm {2C}}}})}$ and for the symmetry reasons, $u=V/2$.  At the same time the recurrent relations Eqs.\eqref{eq:recurrent1}-\eqref{eq:recurrent2} become invariant under the substitution $I_e(\varepsilon-V)= I_h(\varepsilon)$ and reduce to the relation:
\begin{gather}\label{eq:SNSNS_symmetric}
{\mathcal{R}}(\varepsilon, V)I_{\mathrm e}(\varepsilon) -\rho^{(\triangleright)}(\varepsilon) I_{\mathrm e}(\varepsilon-V)- \rho^{(\triangleleft)}(\varepsilon+ V)I_{\mathrm e}(\varepsilon+V)=n_{\scriptscriptstyle{\mathrm F}}(\varepsilon+V)-n_{\scriptscriptstyle{\mathrm F}}(\varepsilon),
\end{gather}
where
\begin{multline}\label{eq:G_N/G_-}
{\mathcal{R}}(\varepsilon, V)\equiv \mathcal{R}(\varepsilon,V/2, V)-\rho^{(\circ)}(\varepsilon+V/2)=
\\
=R_{\scriptscriptstyle{\mathrm N}}(\varepsilon)+(1/2)\sum_{\alpha=\pm}\{\bar R_{\alpha,\varepsilon}^{({{\scriptscriptstyle{\mathrm {1L}}}})}+ \bar R_{\alpha,\varepsilon+\scriptscriptstyle{\mathrm V}}^{({{\scriptscriptstyle{\mathrm {2R}}}})}\},
\end{multline}
where $R_{\scriptscriptstyle{\mathrm N}}(\varepsilon)=R_1+R_2+\bar R_{-,\varepsilon+\scriptscriptstyle{\mathrm V}/2}^{({{\scriptscriptstyle{\mathrm {1C}}}})}+ \bar R_{-,\varepsilon+\scriptscriptstyle{\mathrm V}/2}^{({{\scriptscriptstyle{\mathrm {2C}}}})}$.

The recurrent relation, Eq.\eqref{eq:SNSNS_symmetric}, is similar to that of a (symmetric) SNS junction, see Ref.\cite{Shumeiko,SNSNS} and Appendix~\ref{sec:Appendix_SNS},
but in our case the normal resistance $R_N(\varepsilon)$ becomes energy dependent~\cite{SNSNS}. In other words, a symmetric SNSNS array
has the same transport properties as a single (symmetric!) SNS junction, but with the energy dependent resistance of the normal layer.
The imbalance resistance $\bar R_{-,\varepsilon}^{\scriptscriptstyle{\mathrm{(1C)}}}$ has singularities at the energy
corresponding to the gap edges of the superconducting island in the center of our SNSNS array.
This is the origin of the subharmonic singularities in the current-voltage characteristics at voltages $2\Delta/V=n/2$, $n=1,2,\ldots$, contrasting
 the ``conventional values'' in an SNS junction determined by the relations $2\Delta/V=n$.  It follows from Eq.\eqref{eq:G_N/G_-}
the unusual subharmonic singularities should disappear if the resistance of the normal layer greatly exceeds the resistance of the SN interfaces.
Then $R_{1}\gg \bar R_{-,\varepsilon+\scriptscriptstyle{\mathrm V}/2}^{({{\scriptscriptstyle{\mathrm {1C}}}})}+
\bar R_{-,\varepsilon+\scriptscriptstyle{\mathrm V}/2}^{({{\scriptscriptstyle{\mathrm {2C}}}})}$
and the central superconducting island of the SNSNS array effectively ``disappears'' and the array completely transforms into a SNS junction~\cite{SNSNS}.

\section{Results and Discussion}

Calculation of the current-voltage characteristics ${\mathcal I}(V)$ requires numerical solving of the recurrent relations, Eqs.~\eqref{eq:recurrent1}-\eqref{eq:recurrent2}.
To accomplish the numerical task, we have developed a computational scheme allowing to bypass instabilities caused by the non-analytic behavior of the spectral currents
 $I_{{\mathrm e}({\mathrm h})}(\varepsilon)$, which poses the major computational challenge.
 The procedure is as follows: first, we fix certain chosen energy $\varepsilon$ and identify the set of energies connected through the equations in the given energy
 interval, solving afterwards the resulting subsystem of equations. We then repeat the procedure, until the required energy resolution of $\delta\varepsilon=10^{-5}\Delta$ is achieved.
 Typically, up to $10^6$ linear equations had to be solved for every given voltage, but the complexity of the coupled subsystem depends on the commensurability of $u$ and $V$.

\begin{figure}[t]
\begin{center}
\includegraphics[width=90mm]{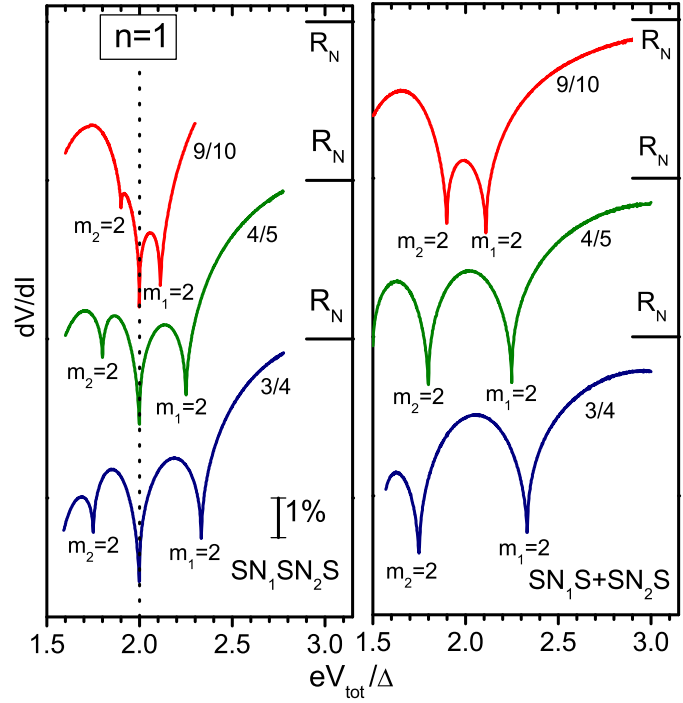}
\caption{Left panel: Differential resistances as functions of the applied voltage $V_{\mathrm{tot}}$ (around $n=1$ in Eq.\,(1)) for the SN$_1$SN$_2$S junction.
The fractions 3/4, and 4/5, and 9/10 represent the ratios of resistances of the normal regions, $R_1/R_2$.
The differential resistance $dV/dI$ of the SN$_1$SN$_2$S junction demonstrates
the pronounced SAT spike at $V_{\mathrm{tot}}=2\Delta/e$, irrespectively to the partial voltage drops.
The SAT spike is sandwiched between the two additional spikes corresponding to individual MAR processes occurring at junctions
SN$_1$S and SN$_2$S for $m_{1}, m_{2}=2$. The voltage positions of these features depend on $R_1/R_2$.
Right panel: The corresponding $dV/dI(V_1+V_2)$ for the two SN$_1$S and SN$_2$S junctions in series as they would have appeared in the absence of the synchronization process,
i.e. in the case where $L_{\scriptscriptstyle{\mathrm C}} > \ell_{\varepsilon}$. These $dV/dI$ dependencies were calculated following~\cite{Shumeiko} (with transmissivity W=1). }
\label{fig:calc}
\end{center}
\end{figure}

Figure~\ref{fig:calc} shows the comparative results for the SNSNS junction and two SNS junctions in series.
The latter corresponds to the case where the size of the central island well exceeds the energy relaxation length, $L_{\scriptscriptstyle{\mathrm C}} > \ell_{\varepsilon}$.
We display the differential resistances as functions of the applied voltage, which demonstrate the singularities in Andreev transmission more profoundly than the $I$-$V$ curves.
There is a pronounced SAT spike in the $dV/dI$ for an SNSNS junction at $V_{\mathrm{tot}}=2\Delta/e$.
The spike appears irrespectively to the partial voltage drops in the normal regions and is absent in the corresponding
curves representing two individual MAR processes at the junctions SN$_1$S and SN$_2$S.

The resonant voltages of the SAT singularities can be found from the  consideration of the quasiparticle trajectories in the space-energy diagrams.
Such a diagram for the first subharmonic, $n=1$ and ratio $R_1/R_2=3/4$ is given in Fig.\,\ref{fig:fig1}.
A quasiparticle starts from the left superconducting electrode with the energy $\varepsilon=-\Delta$ to traverse N$_1$, and the quasiparticle that starts from the central island
 S$_{\mathrm c}$ with the same energy as the incident one to take up upon the current across the island N$_2$, and hit S$_{\scriptscriptstyle{\mathrm R}}$ with the energy
  $\varepsilon=\Delta$ (the ABCD path, the corresponding path for the hole is D$^\prime$C$^\prime$B$^\prime$A$^\prime$).
  In general, relevant trajectories yielding resonant voltages of Eq.\,\eqref{eq:V_position} have the following structure:
  they start and end at the BCS quasiparticle density of states singular points ($\varepsilon=\pm\Delta$), contain the closed polygonal path,
  which include MAR staircases in the normal parts and over-the-gap transmissions and Andreev reflections,
  and pass the density of states singular points at the central island.
  Apart from the main singularities [Eq.~\eqref{eq:V_position}], additional SAT satellite spikes appear at $V=(2\Delta/e)(p+q)/n$,
   where $p/q$ is the irreducible rational approximation
of the real number $r=R_1/R_2$, (we take $R_1<R_2$), and $n\geqslant(p+q)$.

The achieved qualitative understanding enables us to observe that the manifestations of the SAT mechanism in an experimental situation becomes even more pronounced with the growth of the number of SNS junctions in the system.  To see this, let us assume that the resistances of the normal islands in a chain of SNS junctions are randomly scattered around their average value $R_0$ and follow Gaussian statistics with the standard deviation $\sigma_{\scriptscriptstyle{\mathrm R}}=\sigma R_0$, where $\sigma$ is dimensionless. Accordingly, the dispersion of the distribution of the MAR resonant voltages is characterized by the same $\sigma$, and the MAR features get smeared.
Let us distribute the voltage drop $2\Delta/e$ among the $n$ successive islands.
Then the quasiparticle SAT path starts at the lower edge of the superconducting gap at island $j$, traverses $n-1$ intermediate superconducting
islands and hits the edge of the gap at the $j+n$-th island in the chain.
The standard deviation of the voltage drop on the $n$ islands grows as $\sqrt{n}$ resulting in a voltage deviation per one island
$\propto 1/\sqrt{n}$, i.e. the dispersion of the distribution of $V_{\mathrm n}$ drops with increasing $n$: $\sigma_{\scriptscriptstyle{\mathrm{SAT}}}=\sigma/\sqrt{n}$.
 In contrast to the MAR-induced features, with an increase of $n$, the subharmonic spikes at voltages $V_{\mathrm n}$ per junction due to
 SAT processes become more sharp and pronounced.

\section{Conclusions}
In conclusion, we have developed a nonequilibrium theory of the charge transfer across a central
superconducting island in an SNSNS array and found that this island acts as Andreev {\it retransmitter}.
We have shown that the nonequilibrium transport through an SNSNS array is governed
by the synchronized Andreev transmission with the correlated conversion processes
at the opposite NS interfaces of the central island.
The constructed theory is a fundamental building unit for a general quantitative
description of a large  array consisting of many SNS junctions.

\section{Acknowledgments}
We thank A.\,N.~Omelyanchuk for helpful discussions.
The work was supported by the
U.S. Department of Energy Office of Science under the Contract No. DE-AC02-06CH11357, and partually
by the RFBR 10-02-00700, Russian President Science Support foundation mk-7674.2010.2, the Dynasty, Russian  Federal
Programs and the Programs of the Russian  Academy of Science.

\appendix
\section{Recurrent relations for the quasiparticle currents and distribution functions\label{sec:Appendix}}

The Kirchhoff's laws applied for the circuit in Fig.\,\ref{fig:SNSNS}
generate the following linear system of equations:
\begin{eqnarray}\label{eq:system_first}
I_{\mathrm e}(\varepsilon-2V)&=&\frac{\bar n_{{\mathrm h},\varepsilon}^{\scriptscriptstyle{\mathrm {(2R)}}}-
\bar n_{{\mathrm e},\varepsilon-2V}^{\scriptscriptstyle{\mathrm {(2R)}}}}
{\bar R_{{\scriptscriptstyle{\mathrm P}},\varepsilon-V}^{\scriptscriptstyle{\mathrm {(2R)}}}}
+\frac{n_{_\mathrm F}(\varepsilon-V)-
\bar n_{{\mathrm e},\varepsilon-2V}^{\scriptscriptstyle{\mathrm {(2R)}}}}{{\bar R_{+,\varepsilon}^{\scriptscriptstyle{\mathrm {(2R)}}}}},
\\\label{eq:system_second}
I_{\mathrm h}(\varepsilon)&=&\frac{\bar n_{\mathrm h}^{\scriptscriptstyle{\mathrm {(2R)}}}-
\bar n_{{\mathrm e},\varepsilon-2V}^{\scriptscriptstyle{\mathrm {(2R)}}}}
{\bar R_{{\scriptscriptstyle{\mathrm P}},\varepsilon-V}^{\scriptscriptstyle{\mathrm {(2R)}}}}
+\frac{\bar n_{\mathrm h}^{(4)}-n_{_\mathrm F}(\varepsilon-V)}
{\bar R_{+,\varepsilon-V}^{\scriptscriptstyle{\mathrm {(2R)}}}},
\\\label{eq:J_GN2}
I_{\mathrm h}(\varepsilon)&=&\frac{\bar n_{\mathrm h}^{(2)}-
\bar n_{\mathrm h}^{\scriptscriptstyle{\mathrm {(2R)}}}}{R_{2}},
\\\label{eq:A1}
I_{\mathrm h}(\varepsilon)&=&\frac{\bar n_{{\mathrm h},\varepsilon}^{(1)}-
\bar n_{{\mathrm h},\varepsilon}^{(2)}}
{\bar R_{{\scriptscriptstyle{\mathrm D}},\varepsilon-u}}+
\frac{\bar n_{{\mathrm e},\varepsilon-2u}^{(1)}-
\bar n_{{\mathrm e},\varepsilon-2u}^{(2)}}
{\bar R_{{\scriptscriptstyle{\mathrm B}},\varepsilon-u}},
\\\label{eq:A2}
I_{\mathrm e}(\varepsilon-2u)&=&\frac{\bar n_{{\mathrm e},\varepsilon-2u}^{(2)}-
\bar n_{{\mathrm e},\varepsilon-2u}^{(1)}}
{\bar R_{{\scriptscriptstyle{\mathrm D}},\varepsilon-u}}+
\frac{\bar n_{{\mathrm h},\varepsilon}^{(2)}-\bar n_{{\mathrm h},\varepsilon}^{(1)}}
{\bar R_{{\scriptscriptstyle{\mathrm B}},\varepsilon-u}},
\\\label{eq:J_GN1}
I_{\mathrm h}(\varepsilon)&=&\frac{\bar n_{\mathrm h}^{\scriptscriptstyle{\mathrm {(1L)}}}-
\bar n_{\mathrm h}^{\scriptscriptstyle{\mathrm {(1C)}}}}{R_{1}}
\\\label{eq:system_first2}
I_{\mathrm h}(\varepsilon)&=&\frac{\bar n_{\mathrm e}^{\scriptscriptstyle{\mathrm {(1L)}}}-
\bar n_{\mathrm h}^{\scriptscriptstyle{\mathrm {(1L)}}}}
{\bar R_{{\scriptscriptstyle{\mathrm P}},\varepsilon}^{\scriptscriptstyle{\mathrm {(1L)}}}}+
\frac{n_{\scriptscriptstyle{\mathrm F}}(\varepsilon)-\bar n_{\mathrm h}^{\scriptscriptstyle{\mathrm {(1L)}}}}
{\bar R_{+,\varepsilon}^{\scriptscriptstyle{\mathrm {(1L)}}}},
\end{eqnarray}
and
\begin{eqnarray}
I_{\mathrm h}(\varepsilon+2V)&=&\frac{\bar n_{{\mathrm h},\varepsilon+2V}^{\scriptscriptstyle{\mathrm {(2R)}}}-
\bar n_{\mathrm e}^{\scriptscriptstyle{\mathrm {(2R)}}}}
{\bar R_{{\scriptscriptstyle{\mathrm P}},\varepsilon+V}^{\scriptscriptstyle{\mathrm {(2R)}}}}+
\frac{\bar n_{{\mathrm h},\varepsilon+2V}^{\scriptscriptstyle{\mathrm {(2R)}}}
-n_{_\mathrm F}(\varepsilon+V)}
{\bar R_{{\scriptscriptstyle{\mathrm P}},\varepsilon+V}^{\scriptscriptstyle{\mathrm {(2R)}}}},
\\
I_{\mathrm e}(\varepsilon)&=&\frac{\bar n_{{\mathrm h},\varepsilon+2V}^{\scriptscriptstyle{\mathrm {(2R)}}}-
\bar n_{{\mathrm e},\varepsilon}^{\scriptscriptstyle{\mathrm {(2R)}}}}
{\bar R_{{\scriptscriptstyle{\mathrm P}},\varepsilon+V}^{\scriptscriptstyle{\mathrm {(2R)}}}}
+\frac{n_{_\mathrm F}(\varepsilon+V)-\bar n_{{\mathrm e},\varepsilon}^{\scriptscriptstyle{\mathrm {(2R)}}}}
{\bar R_{{\scriptscriptstyle{\mathrm P}},\varepsilon+V}^{\scriptscriptstyle{\mathrm {(2R)}}}},
\\
I_{\mathrm e}(\varepsilon)&=&\frac{\bar n_{{\mathrm e},\varepsilon}^{\scriptscriptstyle{\mathrm {(2R)}}}
-\bar n_{{\mathrm e},\varepsilon}^{\scriptscriptstyle{\mathrm {(2C)}}}}{R_{2}},
\\
I_{\mathrm e}(\varepsilon)&=&\frac{\bar n_{{\mathrm e},\varepsilon}^{\scriptscriptstyle{\mathrm {(2C)}}}
-\bar n_{{\mathrm e},\varepsilon}^{\scriptscriptstyle{\mathrm {(1C)}}}}
{\bar R_{{\scriptscriptstyle{\mathrm D}},\varepsilon+u}^{(12)}}
+\frac{\bar n_{{\mathrm e},\varepsilon+2u}^{\scriptscriptstyle{\mathrm {(2C)}}}
-\bar n_{{\mathrm e},\varepsilon+2u}^{\scriptscriptstyle{\mathrm {(1C)}}}}
{\bar R_{{\scriptscriptstyle{\mathrm B}},\varepsilon+u}^{(12)}},
\\
I_{\mathrm h}(\varepsilon+2u)&=&\frac{\bar n_{{\mathrm h},\varepsilon+2u}^{\scriptscriptstyle{\mathrm {(1C)}}}
-\bar n_{{\mathrm h},\varepsilon+2u}^{\scriptscriptstyle{\mathrm {(2C)}}}}
{\bar R_{{\scriptscriptstyle{\mathrm D}},\varepsilon+u}^{(12)}}
+\frac{\bar n_{{\mathrm e},\varepsilon}^{\scriptscriptstyle{\mathrm {(1C)}}}
-\bar n_{{\mathrm e},\varepsilon}^{\scriptscriptstyle{\mathrm {(2C)}}}}
{\bar R_{{\scriptscriptstyle{\mathrm B}},\varepsilon+u}^{(12)}},
\\\label{eq:1612}
I_{\mathrm e}(\varepsilon)&=&\frac{\bar n_{{\mathrm e},\varepsilon}^{\scriptscriptstyle{\mathrm {(1C)}}}
-\bar n_{{\mathrm e},\varepsilon}^{\scriptscriptstyle{\mathrm {(1L)}}}}{R_{1}},
\\\label{eq:system_last}
I_{\mathrm e}(\varepsilon)&=&\frac{\bar n_{{\mathrm e},\varepsilon}^{\scriptscriptstyle{\mathrm {(1L)}}}
-\bar n_{{\mathrm h},\varepsilon}^{\scriptscriptstyle{\mathrm {(1L)}}}}
{\bar R_{{\scriptscriptstyle{\mathrm P}},\varepsilon}^{\scriptscriptstyle{\mathrm {(1L)}}}}
+\frac{\bar n_{{\mathrm e},\varepsilon}^{\scriptscriptstyle{\mathrm {(1L)}}}
-n_{_\mathrm F}(\varepsilon)}{\bar R_{{\scriptscriptstyle{\mathrm P}},\varepsilon}^{\scriptscriptstyle{\mathrm {(1L)}}}}.
\end{eqnarray}

Eqs.\eqref{eq:system_first}-\eqref{eq:system_last} are
the recurrent relations (i.e. the relations coupling the functions at energy $\varepsilon$ with the functions 
at $\varepsilon\pm V$) for the currents and the distribution functions.

It follows from Eqs.\eqref{eq:J_GN2},\eqref{eq:J_GN1} that
\begin{gather}\label{eq:123}
I_{{\mathrm h},\varepsilon}\,\left[{R_{1}}+{R_{2}}\right]=
\bar n_{\mathrm h}^{\scriptscriptstyle{\mathrm {(2C)}}}-\bar n_{\mathrm h}^{\scriptscriptstyle{\mathrm {(2R)}}}+\bar n_{\mathrm h}^{\scriptscriptstyle{\mathrm {(1L)}}}-\bar n_{\mathrm h}^{\scriptscriptstyle{\mathrm {(1C)}}}.
\end{gather}

The distributions functions entering Eq.\eqref{eq:123} we can express below through the currents.
Combining Eqs.\eqref{eq:A1}-\eqref{eq:A2} we get,
\begin{gather}\label{eq:1999}
\bar n_{\mathrm h}^{({{\scriptscriptstyle{\mathrm {1C}}}})}-\bar n_{\mathrm h}^{({{\scriptscriptstyle{\mathrm {2C}}}})}=
\frac{\left[{I_{{\mathrm h},\varepsilon}\bar R_{{\scriptscriptstyle{\mathrm I}},\varepsilon-u}}
+{I_{{\mathrm e},\varepsilon-2u}\bar R_{{\scriptscriptstyle{\mathrm D}},\varepsilon-u}}\right]
\bar R_{{\scriptscriptstyle{\mathrm D}},\varepsilon-u}\,
\bar R_{{\scriptscriptstyle{\mathrm B}},\varepsilon-u}}
{\left(\bar R_{{\scriptscriptstyle{\mathrm B}},\varepsilon-u}\right)^2
-\left(\bar R_{{\scriptscriptstyle{\mathrm D}},\varepsilon-u}\right)^2}.
\end{gather}
At the same time from Eqs.\eqref{eq:system_first2},\eqref{eq:system_last} follows that
\begin{gather}
\bar n_{{\mathrm h},\varepsilon}^{\scriptscriptstyle{\mathrm {(1L)}}}=
n_{_\mathrm F}(\varepsilon)+\bar R_{+,\varepsilon}^{\scriptscriptstyle{\mathrm {(1L)}}}
\frac{I_{\mathrm e}\,\bar R_{+,\varepsilon}^{\scriptscriptstyle{\mathrm {(1L)}}}-I_{\mathrm h}\,
\left(\bar R_{{\scriptscriptstyle{\mathrm P}},\varepsilon}^{\scriptscriptstyle{\mathrm {(1L)}}}
+\bar R_{+,\varepsilon}^{\scriptscriptstyle{\mathrm {(1L)}}}\right)}
{2\bar R_{+,\varepsilon}^{\scriptscriptstyle{\mathrm {(1L)}}}
+\bar R_{{\scriptscriptstyle{\mathrm P}},\varepsilon}^{\scriptscriptstyle{\mathrm {(1L)}}}},
\end{gather}
and finally from Eqs.\eqref{eq:system_first}-\eqref{eq:system_second} we get
\begin{gather}\label{eq:1812}
\bar n_{{\mathrm h},\varepsilon}^{\scriptscriptstyle{\mathrm {(2R)}}}=n_{\mathrm F}(\varepsilon-V)
+\bar R_{+,\varepsilon-V}^{\scriptscriptstyle{\mathrm {(2R)}}}
\frac{-I_{{\mathrm e},\varepsilon-2V}\,
\bar R_{+,\varepsilon-V}^{\scriptscriptstyle{\mathrm {(2R)}}}
+I_{\mathrm h}\,\left(\bar R_{+,\varepsilon-V}^{\scriptscriptstyle{\mathrm {(2R)}}}
+\bar R_{{\scriptscriptstyle{\mathrm P}},\varepsilon-V}^{\scriptscriptstyle{\mathrm {(2R)}}}\right)}
{2\bar R_{+,\varepsilon-V}^{\scriptscriptstyle{\mathrm {(2R)}}}
+\bar R_{{\scriptscriptstyle{\mathrm P}},\varepsilon-V}^{\scriptscriptstyle{\mathrm {(2R)}}}}.
\end{gather}

Combining Eq.\eqref{eq:123} and Eqs.\eqref{eq:1999}-\eqref{eq:1812}
we find the recurrent relation for the currents, Eq.\eqref{eq:recurrent1}.
Similar procedure helps to derive Eq.\eqref{eq:recurrent2}.

\section{Charge transport in SNS junctions\label{sec:Appendix_SNS}}
We discuss below the transport properties of SNS and SNN' junctions to make a mapping between our technique and the well-known results obtained before us.

The recurrent relations, Eqs.\eqref{eq:recurrent1}-\eqref{eq:recurrent2}, solve the transport problem in a SNS junction in the incoherent regime. Then there is no island, so $\rho^{(\circ)}=0$ and we should remove the island resistances with the indices $(1C)$ and $(2C)$ from the coefficient functions of the recurrent relations. So,
\begin{gather}\label{eq:recurrent1sns}
\mathcal{R}(\varepsilon,- V)I_{\mathrm h}(\varepsilon)-\rho^{(\triangleright)}(\varepsilon)I_{\mathrm e}(\varepsilon)
-\rho^{(\triangleleft)}(\varepsilon -V)I_{\mathrm e}(\varepsilon- V)
=n_{\scriptscriptstyle{\mathrm F}}(\varepsilon)-n_{\scriptscriptstyle{\mathrm F}}(\varepsilon-V),
\\ \label{eq:recurrent2sns}
\mathcal{R}(\varepsilon, V)I_{\mathrm e}(\varepsilon)- \rho^{(\triangleright)}(\varepsilon) I_{\mathrm h}(\varepsilon)- \rho^{(\triangleleft)}(\varepsilon+ V)I_{\mathrm h}(\varepsilon+2V)=n_{\scriptscriptstyle{\mathrm F}}(\varepsilon+V) -n_{\scriptscriptstyle{\mathrm F}}(\varepsilon).
\end{gather}
where, for example,
\begin{gather}\label{eq:a_SNS}
\mathcal{R}(\varepsilon, V)=R_{1}+R_2+(1/2)\sum_{\alpha=\pm} \{\bar R_{\alpha,\varepsilon}^{({{\scriptscriptstyle{\mathrm {1L}}}})}+
  \bar R_{\alpha,\varepsilon+\scriptscriptstyle{\mathrm V}}^{({{\scriptscriptstyle{\mathrm {2R}}}})}\}.
\end{gather}

Eqs.\eqref{eq:recurrent1sns}-\eqref{eq:recurrent2sns} are invariant under the following transformation, $I_{\mathrm e}(\varepsilon-V)\to I_{\mathrm h}(\varepsilon)$,
if \textit{at the same time} we exchange the resistances, $\bar R_{\pm,\varepsilon}^{({{\scriptscriptstyle{\mathrm {1L}}}})}\leftrightarrow \bar R_{\pm,\varepsilon}^{({{\scriptscriptstyle{\mathrm {1R}}}})}$.
Thus the relation, $I_e(\varepsilon-V)=I_h(\varepsilon)$ and the reduction of the recurrent relations to the
one equation for $I_e$ or for $I_h$ as it was done in Ref.\cite{Shumeiko}:
\begin{gather}\label{eq:SNS_symmetric}
{\mathcal{R}}(\varepsilon, V)I_{\mathrm e}(\varepsilon) -\rho^{(\triangleright)}(\varepsilon) I_{\mathrm e}(\varepsilon-V)- \rho^{(\triangleleft)}(\varepsilon+ V)I_{\mathrm e}(\varepsilon+V)=n_F(\varepsilon+V)-n_F(\varepsilon)\,,
\end{gather}
 holds only for a \textit{symmetric} SNS junction with $\bar R_{\pm,\varepsilon}^{({{\scriptscriptstyle{\mathrm {1L}}}})}= \bar R_{\pm,\varepsilon}^{({{\scriptscriptstyle{\mathrm {1R}}}})}$.

To summarize here our consideration [summarized by the recurrent relations Eqs.\eqref{eq:recurrent1sns}-\eqref{eq:recurrent2sns}], reduces to that presented in~ Ref.\cite{Shumeiko} only in the case where the contacts are symmetric and the assumption
$I_{\mathrm{e}}(\varepsilon-V)=I_{\mathrm{h}}(\varepsilon)$ holds.

\printindex

\begin{thebibliography}{99.}

\bibitem{Andreev}
A.\,F.~Andreev,
Zh. Eksp. Teor. Fiz. {\bf 46} (1964) 1823
[Sov. Phys. JETP {\bf 19} (1964) 1228].


\bibitem{SNSexp1}
J.\,M.~Rowell and W.\,E.~Feldmann,
Phys. Rev. \textbf{172}, 393 (1968).

\bibitem{SNSexp2}
P.\,E.~Gregers-Hansen, E.~Hendricks, M.\,T.~Levinsen, and G.\,R.~Pickett,
Phys. Rev. Lett. \textbf{31}, 524 (1973).

\bibitem{SNSexp3}
W.\,M.~van~Huffelen, T.\,M.~Klapwijk, D.\,R.~Heslinga, M.\,J.~de~Boer, and N.~van~der~Post,
Phys. Rev. B \textbf{47}, 5170 (1993).

\bibitem{SNSexp4}
A.\,W.~Kleinsasser, R.\,E.~Miller, W.\,H.~Mallison, and G.\,B.~Arnold,
Phys. Rev. Lett. \textbf{72}, 1738 (1994).

\bibitem{SNSexp5}
E.~Scheer, P.~Joyez, D.~Esteve, C.~Urbina, and M.\,H.~Devoret,
Phys. Rev. Lett. \textbf{78}, 3535 (1997).

\bibitem{SNSexp6}
J.~Kutchinsky, R.~Taboryski, T.~Clausen, C.\,B.~S{\o}rensen, A.~Kristensen,
P.\,E.~Lindelof, J.~Bindslev~Hansen, C.~Schelde~Jacobsen, and J.\,L.~Skov,
Phys. Rev. Lett. \textbf{78}, 931 (1997).

\bibitem{SNSexp7}
A.~Frydman and R.\,C.~Dynes,
Phys. Rev. B \textbf{59}, 8432 (1999).

\bibitem{SNSexp8}
T.~Hoss, C.~Strunk, T.~Nussbaumer, R.~Huber, U.~Staufer, and C.~Sch\"onenberger,
Phys. Rev. B \textbf{62}, 4079 (2000).

\bibitem{SNSexp9}
T.\,I.~Baturina, Z\,.D.~Kvon, R.\,A.~Donaton, M.\,R.~Baklanov, E.\,B.~Olshanetsky,
K.~Maex, A.\,E.~Plotnikov, J.\,C.~Portal,
Physica B \textbf{284}, 1860 (2000).

\bibitem{SNSexp10}
Z.\,D.~Kvon, T.\,I.~Baturina, R.\,A.~Donaton, M.\,R.~Baklanov, K.~Maex,
E.\,B.~Olshanetsky, A.\,E.~Plotnikov, J.\,C.~Portal,
Phys. Rev. B \textbf{61}, 11340 (2000).

\bibitem{MAR1}
T. M. Klapwijk, G. E. Blonder, and M. Tinkham,
Physica B+C (Amsterdam) \textbf{110}, 1657 (1982).

\bibitem{MAR2}
M.~Octavio, M.~Tinkham, G.\,E.~Blonder, and T.\,M.~Klapwijk,
Phys. Rev. B \textbf{27}, 6739 (1983).

\bibitem{MAR3}
K.~Flensberg, J.~Bindslev Hansen, and M.~Octavio,
Phys. Rev. B \textbf{38}, 8707 (1988).

\bibitem{Shumeiko}
E.\,V.~Bezuglyi, E.\,N.~Bratus', V.\,S.~Shumeiko, G.~Wendin, H.~Takayanagi,
Phys. Rev. B \textbf{62}, 14439 (2000). In this work superconductors were considered to be in a local equilibrium and the relation
$I_{\mathrm e}(\varepsilon)=I_{\mathrm h}(\varepsilon-V)$ was satisfied.
This approach was further developed by N.\,M.~Chtchelkatchev, \cite{SNSNS};
however in case of geometrically non-symmetric SNS arrays, it results in an equivalent circuit with
the ennumerable number of elements.

\bibitem{Cuevas}
J.\,C.~Cuevas, J.~Hammer, J.~Kopu, J.\,K.~Viljas, and M.~Eschrig,
Phys. Rev. B \textbf{73}, 184505 (2006).

\bibitem{2DSNS1}
T.\,I.~Baturina, Z.\,D.~Kvon, and A.\,E.~Plotnikov,
Phys. Rev. B \textbf{63}, 180503(R) (2001).

\bibitem{2DSNS2}
T.\,I.~Baturina, Yu.\,A.~Tsaplin, A.\,E.~Plotnikov, and M.\,R.~Baklanov,
JETP Lett. \textbf{81}, 10 (2005).

\bibitem{1DSNS}
T.\,I.~Baturina, D.\,R.~Islamov, and Z.\,D.~Kvon,
JETP Lett. \textbf{75}, 326 (2002).

\bibitem{Fritz}
J.~Fritzsche, R.\,B.\,G.~Kramer, and V.\,V.~Moshchalkov,
Phys. Rev. B \textbf{80}, 094514 (2009).

\bibitem{TiNperf}
T.\,I.~Baturina, A.\,Yu.~Mironov, V.\,M.~Vinokur, N.\,M.~Chtchelkatchev,
A.~Glatz, D.\,A.~Nasimov, A.\,V.~Latyshev,
Physica C (2009) doi:10.1016/j.physc.2009.11.107

\bibitem{SNSNS} N.\,M.~Chtchelkatchev, JETP Lett. \textbf{83}, 250 (2005).

\bibitem{Deutcher}
G.~Deutscher and D.~Feinberg,
Appl. Phys. Lett. \textbf{76}, 487 (2000).

\bibitem{Larkin_Ovchinnikov}
A.I. Larkin and Yu.N. Ovchinnikov,
Sov. Phys. JETP \textbf{41}, 960 (1975); \textit{ibid}, \textbf{46}, 155 (1977).


\bibitem{Kupriyanov-Lukichev}
M. Yu. Kupriyanov and V. F. Lukichev,
Zh. Eksp. Teor. Fiz. \textbf{94}, 139 (1988)
[Sov. Phys. JETP \textbf{67}, 1163 (1988)].

\bibitem{Volkov} A.F. Volkov and T.M. Klapwijk, Phys. Lett. A \textbf{168}, 217 (1992).


\bibitem{Likharev} K.K. Likharev, Rev. Mod. Phys. \textbf{51}, 101 (1979).

\bibitem{Beenakker} C.W.J. Beenakker, Phys. Rev. Lett. \textbf{67}, 3836 (1991).

\bibitem{Tinkham} M.Tinkham, Introduction to superconductivity, Mc.Graw-Hill Inc., 1996.

\bibitem{Chtchelkatchev_spin_filter} N.M. Chtchelkatchev, JETP Lett. \textbf{78}, 230 (2003).

\end{thebibliography}
\end{document}